\documentclass[letter,scriptaddress,twocolumn, prl,showpacs]{revtex4-1}

\usepackage{epsfig}
\usepackage{graphicx}% Include figure files
\usepackage{dcolumn}% Align table columns on decimal point
\usepackage{bm}% bold math
\usepackage{lineno}
\usepackage{amsmath}

\usepackage{hyperref}% add hypertext capabilities

\begin{document}

% Definitions
\def\ca{$\sim$}
\def \EuS {$^{151}$Eu }
\def \EuL{$^{153}$Eu }
\def\Sm {$^{147}$Sm }
\def\Pm{$^{147}$Pm }
\def\LEO{Li$_6$Eu(BO$_3$)$_3$ }

\preprint{APS/123-QED}

\title{Discovery of the \EuS $\alpha$ decay\\}

% Force line breaks with \\
\author{N.~Casali$^{1,2}$}
\author{S.S.~Nagorny$^{1,10}$}
\author{F.~Orio$^{3}$}
\author{L.~Pattavina$^{1}$}
\author{J.W.~Beeman$^{5}$}
\author{F.~Bellini$^{3,4}$}
\author{L.~Cardani$^{3,4}$}
\author{I.~Dafinei$^{3}$}
\author{S.~Di~Domizio$^{6}$}
\author{M.L.~Di~Vacri$^{1}$}
\author{L.~Gironi$^{7,8}$}
\author{M.B.~Kosmyna$^{9}$}
\author{B.P.~Nazarenko$^{9}$}
\author{S.~Nisi$^{1}$}
\author{G.~Pessina$^{8}$}
\author{G.~Piperno$^{3,4}$}
\author{S.~Pirro$^{8}$}
\author{C.~Rusconi$^{8}$}
\author{A.N.~Shekhovtsov$^{9}$}
\author{C.~Tomei$^{3}$}
\author{M.~Vignati$^{3}$}

\affiliation{$^{1}$INFN - Laboratori Nazionali del Gran Sasso, I-67010 Assergi (AQ) - Italy}
\affiliation{$^{2}$Dipartimento di Scienze Fisiche e Chimiche, Universit\`a degli Studi dell'Aquila, I-67100 Coppito (AQ) - Italy}
\affiliation{$^{3}$INFN - Sezione di Roma, I-00185 Roma - Italy}
\affiliation{$^{4}$Dipartimento di Fisica - Universit\`a di Roma La Sapienza, I-00185 Roma - Italy}
\affiliation{$^{5}$Lawrence Berkeley National Laboratory, Berkeley, California 94720, USA}
\affiliation{$^{6}$Dipartimento di Fisica, Universit\`a di Genova and INFN - Sezione di Genova, I-16146 Genova - Italy}
\affiliation{$^{7}$Dipartimento di Fisica, Universit\`a di Milano-Bicocca, Milano I-20126 - Italy}
\affiliation{$^{8}$INFN - Sezione di Milano Bicocca, Milano I-20126 - Italy}
\affiliation{$^{9}$Institute for Single Crystals, National Academy of Sciences of Ukraine, 61001 Kharkov, Ukraine}
\affiliation{$^{10}$Institute for Nuclear Research, National Academy of Sciences of Ukraine, 03680 Kyiv, Ukraine}

\date{\today}% It is always \today, today,
             %  but any date may be explicitly specified

\begin{abstract}
We report on the first compelling observation of $\alpha$ decay of $^{151}$Eu to the ground state of $^{147}$Pm. The measurement was performed using a 6.15~g \LEO crystal operated as a scintillating bolometer. The Q-value and half-life measured are: Q = 1948.9$\pm 6.9(\emph{stat.}) \pm 5.1(\emph{syst.})$~keV, and \mbox{T$_{1/2}=\left( 4.62\pm0.95(\emph{stat.})\pm0.68(\emph{syst.})\right) \times 10^{18}$~y} . The half-life prediction of nuclear theory using the Coulomb and proximity potential model are in good agreement with this experimental result.

\end{abstract}

% PACS, the Physics and Astronomy
                             % Classification Scheme.
%Use showkeys class option if keyword
                              %display desired
\maketitle

%\tableofcontents
\section{INTRODUCTION}

All europium isotopes with mass number ranging from 130 to 153~a.m.u. are potentially unstable nuclides. They can disintegrate through a number of channels, including the $\alpha$ decay~\cite{Q_value}\cite{Q_value1}. However, competing large branching ratio for electron capture and $\beta^-$ decay process make $\alpha$ decay amenable for investigation only for artificially produced $^{147,148,150}$Eu and naturally occurring \EuS  and \EuL.\newline
%just for $^{147,148,150}$Eu and the naturally occurring \EuS  and \EuL isotopes the branching ratio for the $\alpha$ decay is enhanced.\newline

Modern calculations of the partial half-lives of $^{147,148,150,151,153}$Eu mainly follow two approaches, namely the semi-empirical one~\cite{Tav07}, developed in~\cite{Tav07_1} and the Coulomb and proximity potential model (CPPM)~\cite{Santhosh}, proposed in~\cite{Santhosh_08}. The results obtained by the two approaches for the half-life of \EuS range between 8.5$\times$10$^{18}$~y~\cite{Tav07} using semi-empirical calculations and 8.0$\times$10$^{17}$~y~\cite{Santhosh} using CPPM for deformed nuclei (CPPMDN). Given this order of magnitude incompatibility in the theoretical predictions, experimental investigation of such $\alpha$ decays is needed in order to establish which nuclear model better describes these nuclei.\newline

An interesting side-application of observing the $\alpha$ decay of \EuS and measuring its half-life is the impact on the evaluation of promethium concentration in the entire Earth crust~\cite{Promethium}. In fact, one expected promethium production mechanism, besides spontaneous fission of  $^{238}$U, is  $\alpha$ decay of natural \EuS  to the ground state of $^{147}$Pm.

\section{State of the art}
A new era of sensitivity to rare $\alpha$ decays started in 2003 with the application of the scintillating bolometer technique to observe the $\alpha$ decay of $^{209}$Bi with a half-life T$_{1/2} = $1.9$\times$10$^{19}$~y~\cite{nature}. Since then this technique has been successfully exploited to detect $\alpha$ decay of $^{180}$W~\cite{W180}, $\alpha$ decay of $^{209}$Bi to the first excited level~\cite{Bi209-nostro} and to search for $\alpha$ decays of lead isotopes~\cite{PbWO4} --- demonstrating the versatility and power of this technology.\newline

A key advantage of bolometers is the wide choice of detector materials, the detector can often be composed of a significant amount of the nucleus of interest. This results 
in excellent detection efficiency --- the detector is made of the decay source --- which is vital to investigate rare decays. Moreover, high energy resolution (of the order of 
parts per thousand~\cite{reso}) is routinely achieved exploiting this technology. When the bolometer is also an efficient scintillator then the simultaneous readout of the heat and 
light channels enables the identification of the interacting particle nature, i.e whether it is an $\alpha$, $\beta$/$\gamma$, or neutron. This leads to tremendous background suppression, especially for rare $\alpha$ decays with energy transition lower than 2.6~MeV which would otherwise be overwhelmed by the near omnipresent background from $\gamma$ emission of $^{208}$Tl.\newline

Natural europium consists of two isotopes: \EuS (47.81\%) and \EuL (52.19\%)~\cite{ai}. Both are potentially $\alpha$-active with decay energies of Q$_{\alpha}$=1965.0$\pm$1.1~keV and Q$_{\alpha}$=272.4$\pm$2.0~keV~\cite{Q_value}\cite{Q_value1}, respectively. The low Q$_{\alpha}$ for $^{153}$Eu gives no hope for an experimental observation due to the extremely long expected half-life, more than 10$^{140}$~y~\cite{Eu_DAMA}. On the other hand theoretical predictions of the \EuS half-life lie in range of 10$^{17}$-10$^{19}$~y~\cite{Eu_DAMA}, and such sensitivity levels can be achieved by scintillating bolometers.\newline

The $\alpha$ decay of europium isotopes has been recently investigated using a 370~g CaF$_2$:Eu low background scintillator~\cite{Eu_DAMA}, which contained Eu only as a dopant with mass fraction of just 0.5\%. The authors of that paper mention the first indication of the $\alpha$ decay of \EuS nuclide to the ground state of \Pm with a half-life of T$_{1/2}$=5$^{+11}_{-3}\times10^{18}$~y. Nevertheless, due to the large statistical uncertainty, the authors report also a limit on the half-life of the process: T$_{1/2}>$1.7$\times$10$^{18}$~y at 68\% C.L.\newline

In this work we report the first compelling experimental observation of the $\alpha$ decay of \EuS to the ground state of \Pm, using a 6.15~g \LEO crystal (41.1\% of Eu in mass) operated as scintillating bolometer.

\section{Experimental technique}

The crystal was grown using the Czochralski technique in a platinum crucible under air atmosphere in accordance with the thermal conditions described in~\cite{crystal}. The stoichiometric mixture of initial reagents (Li$_2$CO$_3$, Eu$_2$O$_3$ and B$_2$O$_3$) was used to synthesize the \LEO charge by solid phase synthesis as described in~\cite{sintesi}.\newline

The crystal used in this work has  an irregular shape due  to the compromise of having the largest possible mass combined with an acceptable crystal quality. For this reason part of the crystal shows unpolished rounded edges. Moreover, the crystal shows some small cleavage planes as well as some tiny inclusions in part of the structure. Our typical experience with bolometers is that this can negatively affect the energy resolution.\newline

The working principle of a scintillating bolometer consists of measuring the temperature rise produced by the interaction of particles with the crystal absorber, which is operated at a temperature of a few mK. When an ionizing particle deposits energy in the crystal, a large portion of it is converted to heat, while some is converted to scintillation light which is detected by another bolometer facing the crystal. The heat and light signals are both recorded as positive temperature variations by temperature sensors on the main absorber and on the light detector. For the same energy deposition, $\alpha$ and $\beta$/$\gamma$ particles will generate different heat-to-light ratios.\newline

The \LEO crystal was operated in a dilution $^3$He/$^4$He refrigerator in the Gran Sasso underground laboratory of INFN. The crystal was held in position by means of a high-purity copper structure. Germanium Neutron Transmutation Doped thermistors (Ge-NTD) were used as temperature sensors. The Ge-NTDs were coupled to the \LEO crystal and to the light detector (LD) by means of resin epoxy glue. In order to maximize the light collection efficiency, the crystal was surrounded by a reflecting foil (3M VM2002). The operational features of the LD, namely a high purity Ge wafer (diameter 44.5~mm and thickness 300~$\mu$m), are described in~\cite{LD}.

\section{Data analysis}
The detector was operated for 462.2~h. The \LEO crystal and the LD have completely independent read-out. Detailed information on the cryogenic facility electronics, the data acquisition and processing can be found in~\cite{elet_2,elet,acq}.\newline

The amplitude of the heat and light signals is energy-calibrated by means of calibration sources of known energies. The scintillation light is calibrated using a permanent $^{55}$Fe X-ray source facing the LD, we assume a linear dependence of the signal amplitude on the light energy. The heat channel is calibrated using the Q$_{\alpha}$-value of the $\alpha$ decays from internal contamination in the crystal. The calibration curve used is a second order polynomial with zero intercept.\newline

\begin{figure}
\begin{centering}
\includegraphics[width=\columnwidth]{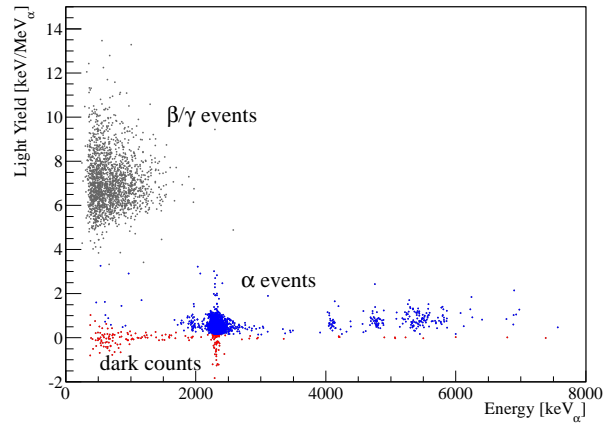}
\caption{Light yield vs heat-energy scatter plot corresponding to 462.2~hours of background measurement with the 6.15~g  \LEO crystal. Three classes of events are identified, each one is characterized by a different Light yield. Colors are used to highlight the $\alpha$ and the dark counts band.}
\label{fig:SP}
\end{centering}
\end{figure}

A useful quantity to consider is the light yield (LY), which we define as the ratio of the energy measured in the light detector in keV to the energy measured in the heat channel of the main absorber in units of MeV. Fig.~\ref{fig:SP} shows LY as a function of the heat-energy measured in the main crystal for each event in the 462.2 hour exposure.  The $\beta$/$\gamma$ and the $\alpha$ interactions in the detector occupy well separated regions of the LY versus heat-energy plan
Ñ in agreement with expectation from BirkÕs law, where $\alpha$ interactions produce lower light signals than those of
$\beta$s/$\gamma$s~\cite{Birks}. In the lower energy region of the scatter plot there is a continuum of events that do not produce light. These so-called dark counts can be induced by several mechanisms. For example, defects such as those present in our crystal can give rise to points in which the scintillation process is strongly reduced. Furthermore, an imperfect crystal in which the structure is stressed by an inhomogeneities can give rise to sudden lattice relaxations, resulting in thermal pulses without the emission of photons. Finally, considering the ratio of the volumes of the Ge-NTD thermistor and the scintillating crystal --- roughly 1:600 --- a fraction of dark counts will arise from interactions that release  energy only inside the thermometer thereby producing a thermal signal without photon emission. In order to reject these events that produce an almost flat background in the region of interest, a cut on the absolute light signal is performed.

\subsection{Alpha energy spectrum}

\begin{figure}[h]
\begin{centering}
\includegraphics[width=\columnwidth]{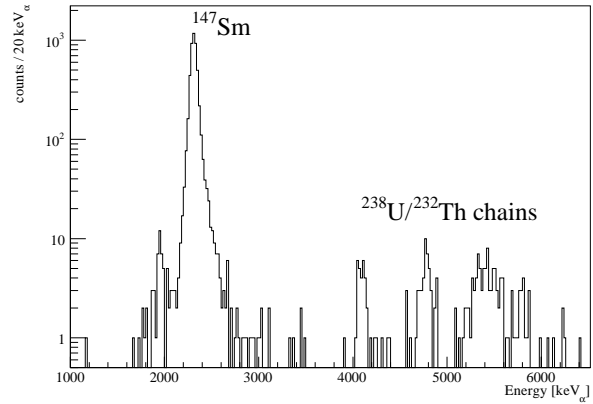}
\caption{Energy spectrum for alpha events selected in the 462.2~h exposure of the 6.15~g \LEO crystal.  The events are attributed to internal contamination of the crystal, the most intense one is $^{147}$Sm $\alpha$ decay at 2311~keV, while the peaks above 4~MeV are due to $\alpha$ decays of daughter nuclides from $^{238}$U/$^{232}$Th chains.}
\label{fig:AlphaRegion}
\end{centering}
\end{figure}

As is clear from  Fig.~\ref{fig:SP}, it is possible to reject  $\beta$/$\gamma$ events by performing a selection on the heat-to-light ratio. The choice LY~$<3.3$~keV/MeV has rejection efficiency of  (99.99$\pm$0.01)\%. To exclude dark
counts, we select events with detected light higher than 0.4 keV. This cut has an efficiency of (96.8$\pm$0.2)\%. The energy spectrum of the remaining events is shown in Fig.~\ref{fig:AlphaRegion}.
The most prominent peak in the energy spectrum is due to an internal contamination of Sm, namely $^{147}$Sm with an activity of 454~mBq/kg. The events in the high energy region, above 4~MeV, have well understood sources: $^{232}$Th decay daughters present at a level of 3.5~mBq/kg; and $^{238}$U daughters particularly $^{226}$Ra and $^{210}$Po, present at the level of 2.9~mBq/kg and 6.2~mBq/kg, respectively. The most intense peaks namely $^{147}$Sm (2311~keV), $^{232}$Th (4083~keV) and $^{226}$Ra (4871~keV) were used for the calibration of the $\alpha$ energy scale. The error on the energy-calibration function at 2311~keV is less than 1~keV.\newline
%\begin{figure}[h]
%\begin{centering}
%\includegraphics[width=\columnwidth]{image/HistoAlpha_zoom.eps}
%\caption{Zoom in of the $\alpha$ spectrum in proximity of the region of interest. The peak in the energy spectrum at 2311~keV corresponds to $^{147}$Sm $\alpha$ decays. There is also a clear peak in the energy window of 1900-2000~keV which corresponds to the $\alpha$ decay of \EuS with expected Q-value at 1964.9$\pm$1.1~keV.}
%\label{fig:AlphaRegionZoom}
%\end{centering}
%\end{figure}

Looking at the spectrum in Fig.~\ref{fig:AlphaRegion}, to the left side of the $^{147}$Sm line we observe a clear peak in the 1900-2000~keV window. There are three known $\alpha$ active nuclides with energy transitions that would reconstruct close to the position of the observed peak. They are $^{144}$Nd (Q$=1905\pm1$~keV, T$_{1/2}$~=~$2.3\times10^{15}$~y), $^{148}$Sm (Q$=1986\pm1$~keV, T$_{1/2}$~=~$7.0\times10^{15}$~y) and $^{152}$Gd (Q$=2203.0\pm1.4$~keV, T$_{1/2}$~=~$1.1\times10^{14}$~y). The concentration of these elements in the crystal was determined using a High Resolution Inductively Coupled Plasma-Mass Spectrometric (HR-ICP-MS) analysis. The obtained results are $4.9\times10^{-6}$~g/g, $4.7\times10^{-6}$~g/g and $91.3\times10^{-6}$~g/g for elemental Nd, Sm and Gd. Thus the contribution of the afore mentioned isotopes to the peak are independently constrained to be only 0.5, 0.1 and 1.5 events.\newline
The robustness of the HR-ICP-MS results is further confirmed by the fact that the measured concentration of the elemental Sm, scaled for the natural isotopic abundance of $^{147}$Sm, agrees with the activity measured with our spectrum analysis within an the error of 15\%.

Given that known $\alpha$ decays are independently constrained to contribute negligibly to this peak, we interpret it as being due to the $\alpha$ decay of $^{151}$Eu.

\subsection{Response function and discovery significance}
\begin{figure}[h]
\begin{centering}
\includegraphics[width=\columnwidth]{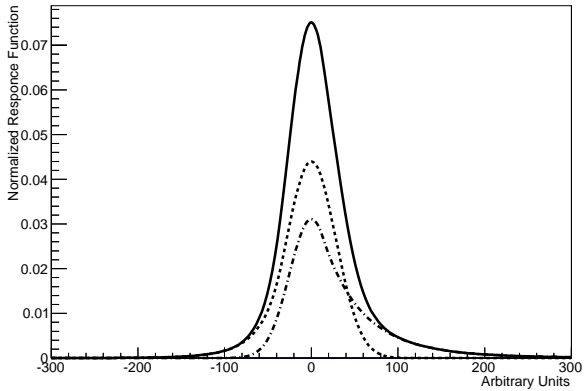}
\caption{Normalized response function of the detector used for $\alpha$ particles interacting in the \LEO crystal. The two dashed lines represent the left and right crystal ball functions, while the solid line is their sum.}
\label{fig:respfunc}
\end{centering}
\end{figure}

We use a crystal ball (CB) function to fit the response of the \LEO bolometer to $\alpha$ particles. We find that the response function (RF) that best describes the $^{147}$Sm peak is the sum of two crystal balls ~\cite{CrystalBall} with the same mean (Q-value), one with a left tail and one with a right tail: \mbox{RF$(Q,E,\delta) = N \cdot [CB_{left}(Q,E) + \delta\cdot CB_{right}(Q,E)]$}.  Here N is an arbitrary normalisation and $\delta$ is the ratio between the two CB functions.
The shape of the response function evaluated on the $^{147}$Sm peak is shown in Fig.~\ref{fig:respfunc}.\newline

With this response function of the bolometer for $\alpha$ particles we perform an unbinned, extended likelihood fit to the $\alpha$ energy spectrum in order to evaluate the number of $^{151}$Eu events. The fit function (FF) used is:
\begin{equation}
\rm{FF(E)} = N_{Sm}\cdot RF(Q_{Sm},E,\delta)+ N_{Eu}\cdot RF(Q_{Eu},E,\delta)\,.
\label{eq:fitfunc}
\end{equation}
$\rm{N_{Sm}}$ and $\rm{N_{Eu}}$ are the number of events in the $^{147}$Sm and $^{151}$Eu peaks and are free parameters of the fit. 
The $^{147}$Sm Q-value is constrained to its established value~\cite{Q_value}\cite{Q_value1}, while the Q-value of \EuS is left as a free parameter.\newline

\begin{figure}[h]
\begin{centering}
\includegraphics[width=\columnwidth]{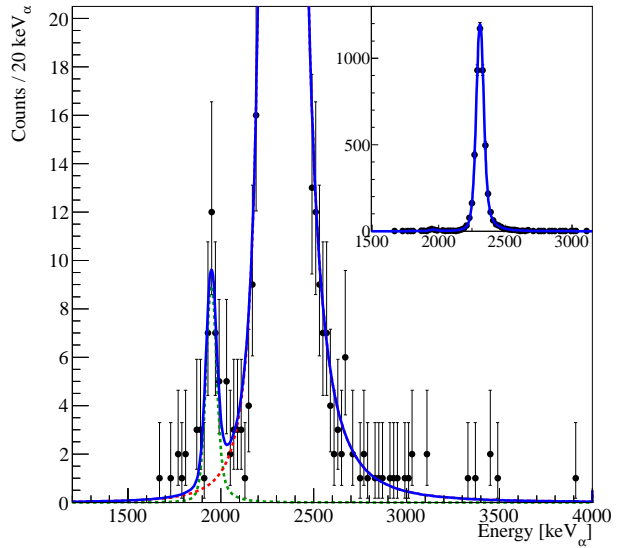}
\caption{Zoom of the $\alpha$ energy spectrum around the region of interest.  Two peaks are visible: $^{147}$Sm at 2311~keV and one at 1949~keV, interpreted as \EuS.  In the inset the full y-scale plot is shown. The response function of the detector is assumed to be the sum of two crystal ball functions and this is used for fitting the two observed peaks in the $\alpha$ energy spectrum. The dashed lines represent the detector response function, while the solid line is the sum of the response function of the two peaks.}
\label{fig:fit}
\end{centering}
\end{figure}

In Fig.~\ref{fig:fit}, the best-fit $\alpha$ energy spectrum from 1.2~MeV to 4~MeV is shown.
The best-fit number of $^{151}$Eu events ($\rm{N_{Eu}}$) corrected for the event selection efficiency, evaluated on the $^{147}$Sm peak to be  (96.8$\pm$0.2)\%, is equal to $37.6\pm7.5$~counts. The best-fit Q-value for the \EuS $\alpha$ transition is 1948.9$\pm 6.9(\emph{stat.})$~keV, and the FWHM energy resolution is 65$\pm$7~keV.\newline

In order to quantify the statistical significance of the excess over the background-only expectation from $^{147}$Sm we adopt the same test used in~\cite{Higgs}. Following that reference, we define
\begin{equation}
q_0 = -2\cdot\rm{ln}\frac{L(\rm{data}\mid bkg(\hat{\theta}_0))}{L(\rm{data}\mid \hat{\mu}\cdot signal(\hat{\theta})+bkg(\hat{\theta}))}\,,
\end{equation}
where $\hat{\theta}_0$ is the set of fit parameter values that maximize the likelihood in the background-only hypothesis, $\hat{\theta}$ is the set of fit parameter values that maximize the likelihood in the background plus signal hypothesis,  and $\hat{\mu}$ is the number of signal events that maximizes the likelihood.  Cases where a fit finds a positive signal, i.e. $\mu >$ 0, correspond to positive values of $q_0$, while in the absence of a signal,  $\mu$ = 0, q$_0$ is zero.  For our fit, the excess of events that we observed gives $q_0 =54$, corresponding to a 7.4~$\sigma$ statistical significance.  Therefore the probability that the measured excess of events is produced by a background fluctuation is of the order of 10$^{-14}$.\newline

\section{$^{151}${Eu} half-life}
The number of \EuS atoms in the 6.15~g \LEO crystal is (4.76$\pm$0.07)$\times 10^{21}$, evaluated using the isotope composition of Eu precisely measured with HR-ICP-MS (47.6$\pm$0.7)\%. %This measurement allows to minimize the systematic uncertainty in the evaluation of the \EuS half-life value. 

To estimate the systematic uncertainty from the choice of fit interval and fit function, we varied both the fitting interval and response function used. For example, we find shrinking the fit range from (1200-4000~keV) to (1800-2800~keV) reduces the best-fit number of \EuS events by 5.3~counts, while changing the response function to a CB plus a gaussian increases the number of \EuS events by 0.9~counts. We include these effects in the systematic uncertainty. Finally, we do not consider the systematic uncertainty arising from the small calibration error  ($<$ 0.1\% at 2311~keV) because it is negligible compared to the detector energy resolution.\newline

Considering the best-fit number of \EuS events corrected for the event selection efficiency (37.6$\pm$7.5~counts), the containment efficiency of \EuS $\alpha$ decay evaluated using Monte Carlo simulation ($\epsilon =$~99.98\%), and the live time of the measurement (462.2~h), we obtain the following value for the half-life for $\alpha$ decay of \EuS to the ground state of {\Pm}:
\begin{equation}
T_{1/2} 
=\left( 4.62\pm0.95(\emph{stat.})\pm0.68(\emph{syst.})\right) \times 10^{18}~y.
\end{equation}

This value is in agreement with half-life interpreted from the indication of \EuS $\alpha$ decay reported in~\cite{Eu_DAMA}. Our measurement concurs, within 1~$\sigma$ with the theoretical prediction using the Coulomb and proximity potential model reported in~\cite{Santhosh},  and is not compatible with the semi-empirical calculation reported in~\cite{Tav07,Poe83,Poe85}.

\section{Conclusion} 
In this work we reported the discovery of \EuS $\alpha$ decay to ground state of \Pm with with a statistical significance of 7.4~$\sigma$. The measured half-life is T$_{1/2}=\left( 4.62\pm0.95(\emph{stat.})\pm0.68(\emph{syst.})\right) \times 10^{18}$~y .\newline

Our result is extremely relevant to identify the best-performing nuclear models for favoured $\alpha$ decays of odd-mass nuclei. In fact, among the different nuclear models previously mentioned in this work, the one in best agreement with our measurement is the Coulomb and proximity potential model (CPPM), where the penetrability potential was evaluated both from the internal and external parts of the nuclear barrier. The computed theoretical value for \EuS half-life is: $4.8 \times 10^{18}$~y, which is less than 1~$\sigma$ away from our experimental value. \newline

The measured Q-value for the decay is 1948.9$\pm6.9(\emph{stat.})\pm 5.1(\emph{syst.})$~keV, in which the statistical error is caused by the small \EuS statistics , while the systematic by varying both the fitting interval and response function used. Our result is compatible with the one reported in \cite{Q_value}\cite{Q_value1} within 2~$\sigma$ interval.
In order to better understand and reduce this discrepancy, further measurements with larger source mass and larger statistics are needed.\newline
%This result was obtained using a Eu-based crystal, such as Li$_6$Eu(BO$_3$)$_3$, operated as scintillating bolometer.\newline

The discovery of \EuS $\alpha$ decay to the ground state of $^{147}$Pm also plays a relevant role in the evaluation of the overall mass of promethium in the Earth's crust. Following the calculation reported in \cite{Eu_DAMA}, we compute the total mass of $^{147}$Pm in the Earth crust to be about 565~g.%, in which the contribution given by the \EuS decay is 12.7$\pm$4.1~g.\newline

%Given the concentration of $^{147}$Pm in uranium ores~\cite{Pm_mines} and the concentration of uranium in the Earth crust~\cite{Eu_DAMA}, we can estimate the total mass of $^{147}$Pm in the Earth crust produced by $^{238}$U spontaneous fission: 552~g. Moreover, if we consider the concentration of natural europium in the Earth crust~\cite{Eu_DAMA}, the half-life of  $^{147}$Pm~\cite{Q_value}\cite{Q_value1} and the \EuS half-life reported in this work, we can evaluate the mass of $^{147}$Pm in equilibrium conditions: 12.7$\pm$4.1~g. Finally, we compute the total mass of $^{147}$Pm in the Earth crust to be about 565~g. \newline

% and it proves the potentialities of the bolometric technique for the study of rare nuclear processes. %We also hope that the half-life value precisely determined in our experiment will motivate theoreticians to develop more simple and complete model for the alpha decay for better description of processes with extremely long half-life. 

\section*{Acknowledgements}
This project was supported by the Italian Ministry of Research under the PRIN 2010ZXAZK9 2010-2011 grant.
This work was also supported by the ISOTTA project, funded within the ASPERA 2nd Common Call for R\&D Activities. Part of the work was carried out thanks to LUCIFER Project, funded  by the European Research Council (FP7/2007-2013) grant agreement no. 247115.\newline

We would like to express special gratitude to Thomas O'Donnell for his careful and accurate revision of the manuscript. Thanks are also due to the LNGS mechanical workshop and in particular to E.~Tatananni, A.~Rotilio, A.~Corsi, and B.~Romualdi for continuous and constructive help in the overall set-up construction. Finally, we are especially grateful to M.~Perego and M.~Guetti for their invaluable help.

%\bibliography{LEO}

\end{document}